\documentclass[10pt,conference]{IEEEtran}
\IEEEoverridecommandlockouts

\usepackage{cite}
\usepackage{tabularx}
\usepackage{amsmath,amssymb,amsfonts}
\usepackage{amsmath, amsthm, amssymb}
\usepackage{graphicx}
\usepackage{textcomp}
\usepackage{enumitem}
\usepackage[table]{xcolor}
\usepackage{bm}
\usepackage{ amsmath }
\usepackage{ amssymb }     
\usepackage{ stmaryrd }
\usepackage{ dsfont }
\usepackage{array} 
\usepackage{booktabs}
\usepackage{bbm}
\usepackage{mathrsfs}
\usepackage{pifont}
\usepackage[normalem]{ulem}
\usepackage{geometry}

\usepackage{algorithm}
\usepackage{algpseudocode}
\usepackage{subcaption}
\usepackage{graphicx}
\usepackage{subcaption}

\begin{document}
\title{Self-Supervised Passive WiFi CSI-Based Motion Tracking with Physics-Informed Deep SSM}%
%Physics-informed Deep state space Modeling for WiFi CSI-Based Motion Estimation

\author{\IEEEauthorblockN{Yifan Feng\IEEEauthorrefmark{1}, Haiyao Yu\IEEEauthorrefmark{1}, Geng Wang\IEEEauthorrefmark{1}, Yunkai Hu\IEEEauthorrefmark{1}, Shenghong Li\IEEEauthorrefmark{2}, Peng Cheng\IEEEauthorrefmark{3}, \\Branka Vucetic\IEEEauthorrefmark{1}, and Yonghui Li\IEEEauthorrefmark{1}}%

\IEEEauthorblockA{\IEEEauthorrefmark{1}School of Electrical and Computing Engineering, The University of Sydney, Australia \\ \IEEEauthorrefmark{2}Data 61, CSIRO, Australia \\ \IEEEauthorrefmark{3}Department of Computer Science and Information Technology, La Trobe University, Australia\\  Email: yifan.feng@sydney.edu.au}}

\maketitle
\begin{abstract}
%Inferring target position from WiFi channel state information (CSI) is challenging due to noisy and heterogeneous multi-domain observations. Measurements such as angle, delay, and Doppler provide only indirect and imperfect information about the underlying state{\color{red}what is the underlying state}, especially under dynamic multipath environments.
%In this paper, we propose a physics-informed Bayesian state space framework for motion estimation from CSI. The target motion is modeled as a latent state, while multi-domain measurements are treated as indirect observations{\color{red}indirect? what about direct observations}. A context-conditioned deep state space model is introduced to capture temporal dependencies and adaptively learn process and observation uncertainties.
%By incorporating geometric constraints, different sensing modalities jointly regulate the consistency and confidence of the inference process. Experimental results demonstrate improved stability{\color{red}how is stability reflected} and accuracy over other existing methods.
Inferring target position from WiFi channel state information (CSI) is challenging due to the noisy and heterogeneous nature of CSI-derived observations.
Measurements such as angle, delay, and Doppler provide only imperfect information about target position and velocity, particularly in dynamic multipath environments where observation reliability varies.
Motivated by this challenge, this paper proposes a physics-informed Bayesian state space framework for passive CSI-based tracking. 
We introduce a context-conditioned deep state space model (DSSM) that captures temporal motion and environmental dynamics, while adaptively modeling time-varying observation uncertainties. 
By incorporating geometric constraints, this framework enables heterogeneous CSI measurement modalities to jointly regularize the consistency and confidence of the tracking process.
Experimental results demonstrate that the proposed framework yields smoother trajectory estimates with reduced temporal fluctuations and outperforms existing methods in tracking accuracy.
\end{abstract}

%%%%%%%%%%%%%%%%%%%%%%%%
%\vspace{-5pt}
\section{Introduction} 
\label{sec:I_introduction}
With the rapid proliferation of Internet-of-Things (IoT) technologies and the widespread deployment of commodity WiFi infrastructure, WiFi channel state information (CSI) sensing has emerged as a key enabler for device-free perception~\cite{11421930}. By exploiting fine-grained variations in wireless channels, CSI enables a wide range of applications, including indoor localization, human tracking, and contactless activity sensing~\cite{indoortrack}. 

WiFi CSI naturally encodes propagation information across multiple domains. Spatial phase differences across antennas provide angular information, frequency-dependent phase variations across subcarriers reflect propagation delay, and temporal phase evolution reveals Doppler shifts induced by motion. These complementary observations characterize the underlying propagation geometry and provide the basis for target tracking from wireless signals.
Prior studies extract angle of arrival (AoA), time of flight (ToF) and Doppler frequency shift from CSI using classical signal processing methods, including subspace-based MUSIC~\cite{schmidt1986music}, maximum-likelihood-based expectation–maximization (EM)~\cite{dempster1977em}, and related variants~\cite{10.1145/3210240.3210314,10.1145/2971648.2971665}. 
%propagation paths and estimate angle, delay, and Doppler. By leveraging signal diversity across the spatial, frequency, and temporal domains, these methods estimate the target motion using geometric relationships between the observations and the location.
%\gray{achieve effective multipath resolution and provide informative geometric cues, which have been widely used for localization and tracking tasks.}
However, the extracted AoA, ToF, and Doppler measurements are inherently imperfect and exhibit distinct limitations~\cite{limit3}. 
AoA estimation is constrained by limited antenna aperture and highly sensitive to multipath interference, which may fluctuate abruptly when the dominant propagation path changes. 
ToF estimation is affected by limited bandwidth and hardware-induced phase distortions, resulting in biased and low-resolution delay estimates. 
Doppler is often more stable, but may still contain distortions or discontinuities in dynamic environments. 
These modality-specific errors significantly degrade localization accuracy and are difficult to eliminate at the raw-signal level due to hardware constraints and complex propagation effects~\cite{9679386}.
This motivates robust tracking strategies that explicitly account for measurement uncertainty after CSI feature extraction.
%{\color{red}need some transition sentences say that these limitations cannot be solved at raw signal level so we propose algorithms at measurement level}
    
To improve tracking accuracy, existing works commonly incorporate temporal tracking using Bayesian filters, such as the Kalman filter (KF)~\cite{kalman}, extended Kalman filter (EKF)~\cite{barshalom2011tracking}, and particle filter (PF)~\cite{gustafsson2010pf}.
Other learning-based approaches are also adopted to capture more complex motion patterns~\cite{deepfi2015}.
However, accurate passive CSI tracking remains challenging in dynamic multipath environments, where the target carries no device and its position can only be inferred indirectly through reflected paths. 
In practice, the reliability of CSI observations is strongly time-varying and depends on propagation conditions induced by target motion and environmental dynamics. 
Such non-stationary observation behaviors are difficult to represent using predefined linear or parametric models. 
Consequently, conventional filters with fixed noise assumptions cannot adapt to changing propagation conditions. 
Although learning-based approaches can improve modeling capability, they require labeled data and generalize poorly to unseen environments~\cite{limit2}. These limitations highlight the need for a framework that integrates physical geometric modeling with adaptive characterization of measurement uncertainties.

%\red{why differ in nature, provide more explanations, such as are time-varying and differ in nature since the dynamic environment will change when people are moving, how to model.... }

%\red{what is greater flexibility? You mean generalization?, i suggest change to more accurate} 

%\red{why limitations motivate the need for a unified framework that can jointly model physical constraints, temporal dependencies, and time-varying observation reliability? SSM also uses the physical constraints and temporal dependencies. The reader only sees from the above description that we need a more accurate and more generalized model.}

%\red{rewrite the above three paragraphs. In the first paragraph, it is mentioned that existing work, e.g., music, extracts Doppler, AoA, and ToF, and uses geometric information to localize the target. The second paragraph talks about the limitations of AoA, tof, and Doppler, and mentions existing work uses kf,ekf, and pf to mitigate the error, or uses a learning-based model. The third paragraph mentioned the limitation of SSM and learning-based models, such that it works; however, it can not effectively track the target since the multipath, dynamic environment..., which is hard to model. learning-based requires label, and low generalization.... then provide that To address these challenges, we proposed a self-supervised DSSM...  }

In this paper, we propose a physics-informed deep state space model (DSSM) for passive WiFi CSI-based motion tracking. 
Building upon the DSSM formulation~\cite{Wang/TSP/DSSM/2026,dssm}, we integrate domain-specific geometric constraints and data-driven observation modeling to jointly estimate motion states and time-varying observation uncertainty. 
This design provides a unified treatment of physical structure and observation reliability, leading to more robust motion tracking.
The main contributions of this paper are summarized as follows:
%\gray{from CSI. The target motion is modeled as a latent state, while multi-domain measurements, including AoA, ToF, and Doppler frequency shift, are treated as indirect observations related to the state through geometric propagation relationships. This enables different sensing modalities to jointly constrain the estimation process in a unified manner.}

%\gray{To improve robustness, we further introduce a context-conditioned unsupervised deep state space model, where historical observations are summarized into a context variable to adapt the transition and observation uncertainty over time. This allows the model to capture time-varying observation reliability in dynamic environments.}

%\begin{itemize}
   % \item We propose a physics-informed state space framework that links latent motion with multi-domain CSI measurements via geometric relationships.

%\red{you mentioned physics-informed?what the advantage? In the writing contribution section, for each proposal, you need to provide its advantages.  }
 %   \item We develop a context-conditioned deep state space model (DSSM) that adaptively models time-varying uncertainty for robust motion estimation.
  %  \item We establish a fully unsupervised learning framework without requiring labeled trajectory data.
%\end{itemize}

%\red{give more explanation of each item. Contributions 3 should be combined with 1 or 2. add contribution 3 for experimental results.}
%\red{you also need to emphasize the proposed system is an unsupervised learning. }

\begin{itemize}
    \item We propose a physics-informed state space formulation that maps latent motion states to multi-domain CSI measurements via geometric relationships.
    This enables physically consistent and robust tracking without labeled data.
    
    \item We develop a context-conditioned DSSM that adaptively models time-varying and modality-dependent uncertainties. By leveraging historical context, the proposed approach effectively handles non-stationary observation noise and enhances tracking accuracy.
    
        \item We conduct extensive experiments in a practical indoor environment. The results demonstrate that the proposed method reduces localization error by up to 31.9\% and achieves 0.67~m mean absolute error (MAE), outperforming existing baselines.
\end{itemize}

\section{SYSTEM MODEL AND PROBLEM FORMULATION}
\begin{figure}[t]
    \centering
    \includegraphics[width=0.45\textwidth]{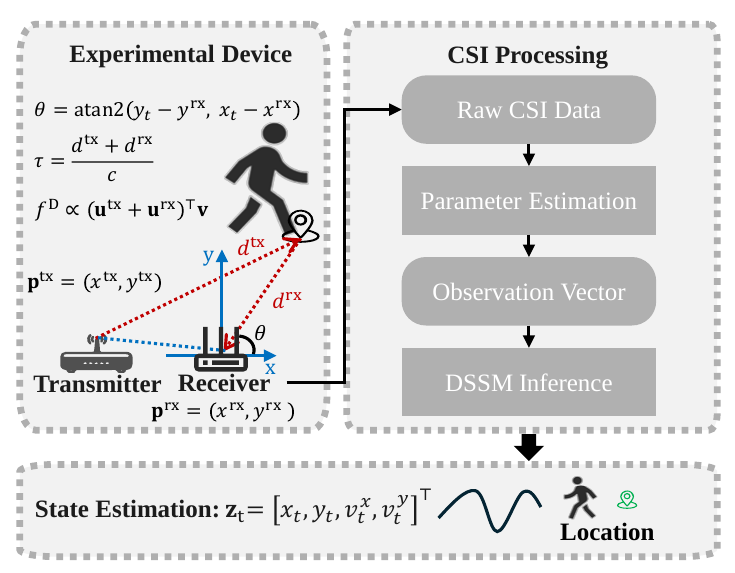}
    \vspace{-0.2cm}
    \caption{Overview of the proposed passive WiFi CSI-based tracking framework.} 
    \label{fig:system}
    \vspace{-0.4cm}
\end{figure}

\subsection{CSI Signal Model}

We consider a commodity WiFi sensing system with a single-antenna transmitter (TX) and a receiver (RX) with $M$ antennas, where the position of a human is estimated from the CSI measured by the receiver, as shown in Fig. 1. The carrier frequency is denoted by $f^{\rm c}$, the frequency of the $k$-th subcarrier is denoted by $f_k$, and the sampling rate of the receiver is denoted by $f^{\rm s}$.
Existing literature assumes that human motion within a short time window can be approximated as constant, as the short interval minimizes velocity variation and allows for a simplified motion model~\cite{9679386}. 
In each window, $M \times K \times S$ CSI samples are collected, where $K$ is the number of subcarriers and $S$ is the number of time slots. The received CSI from the $t$-th window, $m$-th antenna, $k$-th subcarrier, and $s$-th time slot is denoted by $H_{m,k,s}[t]$ as
\begin{equation}\label{csi_model}
\begin{aligned}
H_{m,k,s}[t]
&=
\sum_{\ell=1}^{L} \alpha_{\ell,s}[t] 
\exp\!\Big(
j 2\pi 
\big(
\!-f_k \tau_{\ell}[t] \\
&\quad
- \frac{f_k}{c}(m-1)d \sin \theta_{\ell}[t]
+ f^{\rm D}_{\ell}[t]\, s
\big)
\Big)
\end{aligned}
\end{equation}
where $L$ denotes the number of propagation paths during the $t$-th window, $c$ is the speed of light and $d$ is the inter-antenna spacing. $\alpha_{\ell,s}[t]$, $\tau_{\ell}[t]$, $f^{\rm D}_{\ell}[t]$, and $\theta_{\ell}[t]$ denote the complex gain, propagation delay, Doppler frequency shift, and AoA of the $\ell$-th path, respectively.

Among the $L$ propagation paths, we focus on the dominant dynamic path associated with the target-induced reflection, which carries motion-related information. In practical WiFi sensing systems, the AoA, ToF, and Doppler of this path cannot be directly obtained from \eqref{csi_model}, but are estimated using dedicated parameter estimation methods such as SAGE or WiDFS~\cite{9679386, 10.1145/3210240.3210314}. 
Specifically, at each time window $t$, these methods provide multi-domain observations
\begin{equation}
\mathbf{o}_t =
\left[
\hat{\theta}_t,\;
\hat{\tau}_t,\;
\hat{f}^{\rm D}_t
\right]^\top \in \mathbb{R}^{3\times 1},
\end{equation}
where $\hat{\theta}_t$, $\hat{\tau}_t$, and $\hat{f}^{\rm D}_{t}$ denote the estimated AoA, ToF, and Doppler frequency of the dominant reflected path\footnote{We omit their subscript $\ell$ for notational simplicity}, respectively.
\subsection{Problem Formulation}\label{sec:Problem Formulation}

To describe the underlying motion of the target, we define a target latent state at time window $t$ as
\begin{equation}
\mathbf{z}_t = [\mathbf{p}_t,\mathbf{v}_t]^\top=[x_t, y_t, v^{x}_{t}, v^{y}_{t}]^\top \in \mathbb{R}^{4\times 1},
\end{equation}
where $\mathbf{p}_t=(x_t, y_t)$ and $\mathbf{v}_t=(v^x_{t}, v^y_{t})$ denote the position and velocity of the target, respectively.
The CSI tracking problem is then formulated as a sequential filtering task, where the latent state $\mathbf{z}_t$ is inferred from the observation history $\mathbf{o}_{1:t}$ through the conditional distribution $p(\mathbf{z}_t \mid \mathbf{o}_{1:t})$.

However, the CSI measurements $\mathbf{o}_t$ are time-varying with modality-dependent uncertainties due to factors such as multipath propagation, target motion, and hardware-induced distortions, which are difficult to capture using conventional state space models (SSMs), %such as KF~\cite{kalman}, EKF~\cite{10069890}, and PF~\cite{5546308}, 
thereby degrading tracking accuracy.
%using predefined SSMs, such as Kalman filtering (KF)~\cite{kalman}, extended Kalman filtering (EKF)~\cite{10069890}, and particle filtering (PF)~\cite{5546308}. As a result, the observations are not only noisy but also exhibit time-varying and modality-dependent reliability. Directly inferring the target motion from instantaneous observations is therefore unreliable.
%\gray{To improve robustness against observation noise and temporal inconsistency, existing works typically employ SSMs, such as Kalman filtering (KF)~\cite{kalman}, extended Kalman filtering (EKF)~\cite{10069890}, and particle filtering (PF)~\cite{5546308}, to establish a sequential relationship between observations and the underlying motion state. However, the observation errors across different modalities exhibit distinct characteristics, including time-varying uncertainty, bias, and occasional inconsistencies, which are difficult to capture using a unified linear or parametric model. This mismatch limits the effectiveness of conventional SSM-based methods.}
To address this limitation, we introduce the DSSM framework to model the time-varying reliability of multi-domain observations and improve target state estimation in dynamic multipath environments.
\begin{figure}[ht]
    \centering
    \includegraphics[width=\linewidth]{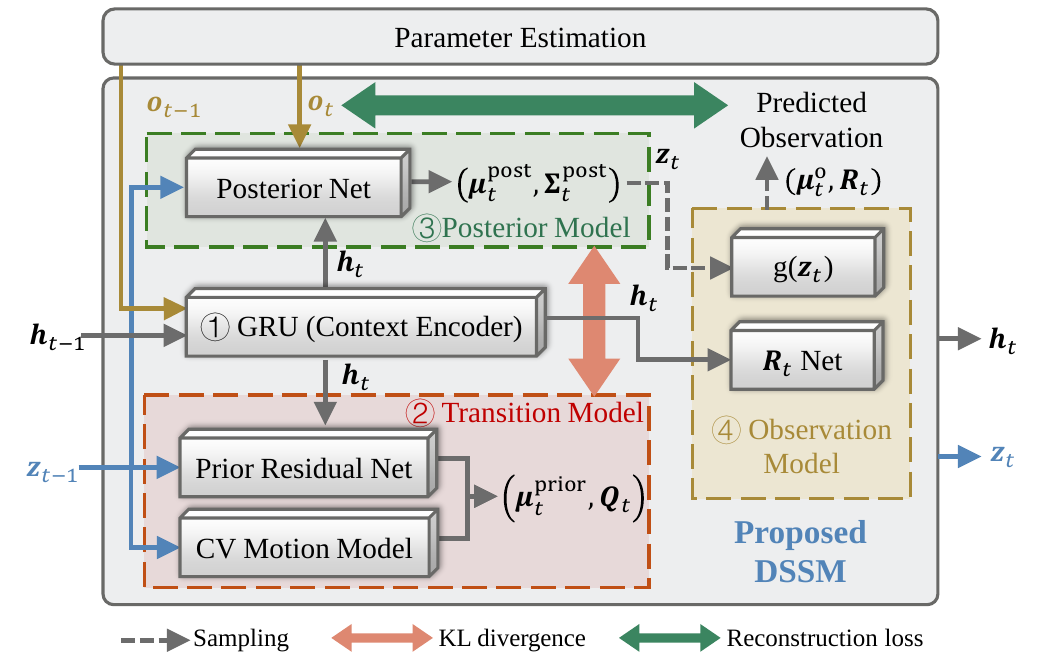}
    \caption{Proposed DSSM framework.} 
    \vspace{-0.5cm}
    \label{fig:placeholder}
\end{figure}

\section{Physics-Informed Deep State Space Model for CSI-Based Motion Estimation}
In this section, we present the proposed physics-informed DSSM for motion estimation from CSI. As shown in Fig.~\ref{fig:placeholder}, the DSSM contains a context encoder, a transition model, a posterior model, and an observation model.  Details of each model and the self-supervised training process will be elaborated in the next subsections.
%We first formulate the problem and then describe the design of model functions, and finally develop the inference scheme for state estimation and model training.
\vspace{-0.4cm}
\subsection{Context Encoder}

As mentioned in Section~\ref{sec:Problem Formulation}, the dynamic multipath propagation and environmental changes cannot be captured by the conventional formulations. To address this limitation, we introduce an auxiliary context variable $\mathbf{h}_t$, which is recursively updated from the previous context $\mathbf{h}_{t-1}$ and historical observations. 
The variable $\mathbf{h}_t$ serves as a compact representation of temporal context that cannot be described by the low-dimensional motion state. 
In particular, it implicitly captures time-varying factors reflected in the observation sequence, such as changes in propagation conditions and variations in observation reliability across different modalities.

%which encodes historical information of the environment and observation process. Specifically, $\mathbf{h}_t$ captures time-varying factors that are not explicitly represented by the motion state, including changes in propagation conditions and variations in observation reliability across different sensing modalities.

%The latent motion state $\mathbf{z}_t$ captures the geometric evolution of the target, but it does not explicitly account for the time-varying uncertainty and reliability of observations. In practical WiFi sensing scenarios, these factors are influenced by dynamic multipath propagation and environmental changes, and cannot be fully represented by the motion state alone. As a result, models that rely solely on $\mathbf{z}_t$ may struggle to adapt to dynamic sensing conditions.

%In CSI-based localization, these effects are closely related to the underlying propagation geometry and signal conditions, and directly influence the consistency and reliability of multi-domain observations. By incorporating $\mathbf{h}_t$, the model is able to implicitly represent such evolving factors and adapt to non-stationary sensing environments.

We model $\mathbf{h}_t$ as the hidden state of a gated recurrent unit (GRU), and each $\mathbf{h}_t$ in the $t$-th time window is computed as
\begin{equation}
\mathbf{h}_t = f_{\theta^{\mathbf h}}(\mathbf{o}_{t-1}, \mathbf{h}_{t-1}) \in \mathbb{R}^{d_h \times 1},
\end{equation}
where $d_h$ denotes the dimension of the hidden state $\mathbf{h}_t$, the initial state is defined as $\mathbf{h}_0 = \mathbf{0} \in \mathbb{R}^{d_h \times 1}$, and $f_{\theta^{\mathbf h}}(\cdot)$ denotes a GRU parameterized by $\theta^{\mathbf h}$. Accordingly, the transition and observation models of the DSSM are augmented by $\mathbf{h}_t$ as
\begin{equation}
\mathbf{z}_t \sim p_{\theta^{\mathbf z}}(\mathbf{z}_t \mid \mathbf{z}_{t-1}, \mathbf{h}_t),
\quad
\mathbf{o}_t \sim g_{\theta^{\mathbf o}}(\mathbf{o}_t \mid \mathbf{z}_t, \mathbf{h}_t),
\end{equation}
where $\theta^{\mathbf z}$ and $\theta^{\mathbf o}$ are the corresponding parameters.% of the , and the observation likelihood $p_{\theta^{\mathbf o}}(\mathbf{o}_t \mid \mathbf{z}_t, \mathbf{h}_t)$ is defined as a Gaussian distribution with mean $\mathbf{g}(\mathbf{z}_t)$ and covariance $\mathbf{R}_t$.

\subsection{Physics-Informed Transition and Observation Models}
In this subsection, we design the physics-informed transition and observation models by combining physically derived dynamics with learnable uncertainty modeling.

\subsubsection{Transition Model}

We model $\mathbf{z}_t$ as a Gaussian distribution
\begin{equation}
\mathbf{z}_t \sim 
\mathcal{N}\!\left(
\boldsymbol{\mu}_t^{\mathrm{prior}},
\mathbf{Q}_t
\right),
\end{equation}
where $\boldsymbol{\mu}_t^{\mathrm{prior}}\in\mathbb{R}^{4\times1}$ is the prior mean that characterizes the expected state transition, and the covariance matrix 
$\mathbf{Q}_t=\mathrm{diag}\big((\boldsymbol{\sigma}_t^\mathrm{prior})^2\big)\in\mathbb{R}^{4\times4}$ captures the associated uncertainty.

In CSI-based localization, this uncertainty mainly arises from the stochastic nature of human motion. Specifically, the actual target motion may deviate from the assumed motion model due to non-constant velocity, acceleration, and turning behaviors within the tracking sequence. Such deviations introduce uncertainty in the state evolution, which is captured by $\mathbf{Q}_t$.
%{\color{red}i dont understand how the second point is related to Q? move to observation model?}.
%{\color{red}im not sure what this means, do you mean that the actual motion deviates from the assumed motion? and also why multipath affects motion?}.

\underline{Prior mean $\boldsymbol{\mu}_t^{\mathrm{prior}}$}: The prior mean is constructed based on a physics-based motion approximation, denoted by $\boldsymbol{\mu}_t^{\mathrm{phy}}$, and a learnable correction, denoted by $\Delta \boldsymbol{\mu}_t$. 
Specifically, 
\begin{equation}
\begin{aligned}
    \boldsymbol{\mu}_t^{\mathrm{prior}}
&=
\boldsymbol{\mu}_t^{\mathrm{phy}}
+
\Delta \boldsymbol{\mu}_t=f^{\mathrm{phy}}(\mathbf{z}_{t-1})+f_{\theta_{\mu}^{\mathbf z}}(\mathbf{z}_{t-1},\mathbf{h}_t),
\end{aligned}
\end{equation}
where $\boldsymbol{\mu}_t^{\mathrm{phy}}$ is determined by a physical motion model and the learnable correction term $\Delta \boldsymbol{\mu}_t$ is obtained via a Multi-Layer Perceptron (MLP) network denoted by $f_{\theta_{\mu}^{\mathbf z}}$ with inputs $\mathbf{z}_{t-1}$ and $\mathbf h_t$ and trainable parameters $\theta_{\mu}^z$. 
For $f^{\mathrm{phy}}(\mathbf{z}_{t-1})$, a constant-velocity (CV) motion model is applied such that
\begin{equation}
\begin{aligned}
\boldsymbol{\mu}_t^{\mathrm{phy}}
=f^{\mathrm{phy}}(\mathbf{z}_{t-1})=\begin{bmatrix}
1 & 0 & \Delta t & 0 \\
0 & 1 & 0 & \Delta t \\
0 & 0 & 1 & 0 \\
0 & 0 & 0 & 1
\end{bmatrix}\mathbf{z}_{t-1},
\end{aligned}
\end{equation}
where $\Delta t$ is the time difference between two adjacent time windows. 
The CV model is a widely adopted assumption in target tracking that provides a simple yet effective approximation of motion dynamics over short time windows. In WiFi-based localization scenarios, human motion is typically smooth within a time window, making the CV model sufficient to capture the dominant motion trend. This formulation provides a physically interpretable and stable baseline for modeling the state evolution.

\underline{Prior standard deviation $\boldsymbol{\sigma}_t^{\mathrm{prior}}$}: 
The standard deviation $\boldsymbol{\sigma}_t^{\mathrm{prior}}$ is predicted by the prior residual net as
\begin{equation}
\boldsymbol{\sigma}_t^{\mathrm{prior}}
=
\mathrm{Softplus}\!\left(
f_{\theta_\sigma^{\mathbf{z}}}(\mathbf{z}_{t-1}, \mathbf{h}_t)
\right)
+
\epsilon_Q,
\end{equation}
where $f_{\theta_\sigma^{\mathbf{z}}}$ is a MLP parameterized by $\theta_{\sigma}^{\mathbf z}$ with inputs $\mathbf{z}_{t-1}$ and $\mathbf{h}_t$. The constant $\epsilon_Q > 0$ is introduced to ensure numerical stability, while the Softplus function guarantees that the standard deviation remains strictly positive.

The formulation of the state transition parameters $\theta^{\mathbf{z}}=\{\theta^{\mathbf z}_{\mu},\theta^{\mathbf z}_{\sigma}\}$ enables the model to adaptively adjust the confidence of the transition according to time-varying dynamics, while preserving the interpretability of the underlying physical motion model.

% \red{rewrite similarly, first provide the \begin{equation}
% p_{\theta}(\mathbf{o}_t \mid \mathbf{z}_t, \mathbf{h}_t)
% =
% \mathcal{N}\!\left(
% \mathbf{o}_t;
% \mathbf{g}(\mathbf{z}_t),
% \mathbf{R}_t
% \right),
% \end{equation}, then introduce the $\mu$, and $\delta$, respectively.  }
%We model the observation likelihood of $\mathbf{o}_t$ as a Gaussian distribution
%\begin{equation}
%\mathbf{o}_t
%\sim
%\mathcal{N}\!\left(
%\mathbf{g}(\mathbf{z}_t),
%\mathbf{R}_t
%\right),
%\end{equation}
%{\color{green}inconsistent with eq 10}
%where the mean value $\mathbf{g}(\mathbf{z}_t)\in\mathbb{R}^{3\times1}$ {\color{green} i suggest introduce another symbol $\mu$ here as the output of $g$} is the deterministic observation function determined by the geometric projection between motion state and observations.  $\mathbf{R}_t=\mathrm{diag}\left((\boldsymbol{\sigma}_t^{\mathrm{o}})^2 \right) \in \mathbb{R}^{3\times3}$ denotes the time-varying observation noise covariance matrix, computed through a neural network, to capture the time-varying and modality-dependent observation uncertainty.
\subsubsection{Observation Model}

We model the observation likelihood as a Gaussian distribution
\begin{equation}
\mathbf{o}_t
\sim
\mathcal{N}\!\left(
\boldsymbol{\mu}_t^{\mathrm{o}},
\mathbf{R}_t
\right),
\end{equation}
where $\boldsymbol{\mu}_t^{\mathrm{o}} = {g}(\mathbf{z}_t)$ denotes the expected observation given the posterior latent state $\mathbf{z}_t$, which is sampled from the posterior distribution as detailed in Section~\ref{sec:3c}. 
The physics-informed observation model ${g}(\cdot)$ is defined by the geometric projection from the motion state to the observation space, and $\mathbf{R}_t = \mathrm{diag}\!\left( (\boldsymbol{\sigma}_t^{\mathrm{o}})^2 \right) \in \mathbb{R}^{3\times3}$ is the observation covariance. The observation uncertainty arises from imperfections in CSI-derived measurements, including multipath propagation, path switching, non-line-of-sight (NLoS) effects, limited bandwidth, and hardware-induced distortions, which degrade the reliability of the observations $\mathbf{o}_t$. These effects are captured by the adaptive modeling of $\mathbf{R}_t$.

\underline{Observation mean $\boldsymbol{\mu}_t^{\mathrm{o}}$}: The observation mean value is defined as
\begin{equation}
{g}(\mathbf{z}_t)
=[g_{\tau}(\mathbf{z}_t),
g_{\theta}(\mathbf{z}_t),
g_{D}(\mathbf{z}_t)]^{\top},
\end{equation}
where $g_{\tau}(\mathbf{z}_t)$, $g_{\theta}(\mathbf{z}_t)$, and $g_{D}(\mathbf{z}_t)$ are the reconstructed ToF, AoA, and Doppler estimated under the propagation geometric relation between TX, RX%{\color{red}defined anywhere before?}
, and the prior motion state, respectively. Specifically,
\begin{itemize}
    \item The reconstructed ToF is determined by the total dynamic path, as shown in Fig.~\ref{fig:system}, given by
    \begin{equation}
    g_{\tau}(\mathbf{z}_t)=\frac{d^{\mathrm{tx}}+d^{\mathrm{rx}}}{c}
    =\frac{\|\mathbf{p}_t - \mathbf{p}^\mathrm{tx}\|+\|\mathbf{p}_t - \mathbf{p}^\mathrm{rx}\|}{c},
    \end{equation}
where $\mathbf{p}^{\mathrm{tx}}=(x^{\mathrm{tx}},y^{\mathrm{tx}})$ and $\mathbf{p}^{\mathrm{rx}}=(x^{\mathrm{rx}},y^{\mathrm{rx}})$ are the locations of transmitter and receiver, respectively.
\item The reconstructed AoA is determined as
\begin{equation}
g_{\theta}(\mathbf{z}_t)
= \operatorname{atan2}(y_t - y^{\mathrm{rx}},\; x_t - x^{\mathrm{rx}}).
\end{equation}
\item The reconstructed Doppler is determined by projecting the target velocity in two directions from the target to the TX and RX as
\begin{equation}
g_{D}(\mathbf{z}_t)= \frac{f^{\mathrm c}(\mathbf{u}^{\mathrm{tx}}_{t} + \mathbf{u}^{\mathrm{rx}}_{t})^{\top}\mathbf{v}_t}{c},
\end{equation}
where $\mathbf{u}^{\mathrm{tx}}_{t}$ and $\mathbf{u}^{\mathrm{rx}}_{t}$ are the unit vectors from the target toward the TX and RX, respectively, given by
\begin{equation}
\mathbf{u}_t^{\mathrm{tx}}=\frac{\mathbf{p}^{\mathrm{tx}} - \mathbf{p}_t}{\|\mathbf{p}^{\mathrm{tx}} - \mathbf{p}_t\|}, \qquad
\mathbf{u}_t^{\mathrm{rx}}=\frac{\mathbf{p}^{\mathrm{rx}} - \mathbf{p}_t}{\|\mathbf{p}^{\mathrm{rx}} - \mathbf{p}_t\|}.
\end{equation}
    \end{itemize}

\underline{Observation standard deviation $\boldsymbol{\sigma}_t^{\mathrm{o}}$}: 
The standard deviation of the observation $\boldsymbol{\sigma}_t^{\mathrm{o}}$ is estimated by a MLP network with $\mathbf{h}_t$ as input, which serves as the observation noise estimation module corresponding to the $\mathbf{R}_t$ Net in Fig.~2. Specifically  
\begin{equation}
\boldsymbol{\sigma}_t^{\mathrm{o}}
=\mathrm{Softplus}\big(
f_{{\theta}^{\mathbf{o}}}(\mathbf{h}_t)
\big)
+
\epsilon_R,
\end{equation}
where $\theta^{\mathbf{o}}$ denotes the trainable parameters, and $\epsilon_R > 0$ is a small constant. 
% The observation covariance is then constructed as 
% $\mathbf{R}_t = \mathrm{diag}\!\left((\boldsymbol{\sigma}_t^{\mathrm{o}})^2\right)$.
%\red{introduce something about $\epsilon_R$}
%{\color{green}discuss why model this and what this does}
%where $\mathrm{Head}_R(\cdot)$ maps the context variable to modality-wise uncertainty. This formulation enables the model to adaptively adjust the confidence of each observation dimension according to the temporal context encoded in $\mathbf{h}_t$.

\subsection{Deep State Space Model Training}\label{sec:3c}
The training objective of the DSSM is the evidence lower bound (ELBO) derived in~\cite{Wang/TSP/DSSM/2026}
\begin{equation}\label{eq:elbo}
\begin{aligned}
\mathcal{L}
=
\sum_{t=1}^{T}
\Big[
&\mathbb{E}_{q_{\phi}(\mathbf{z}_t \mid \mathbf{z}_{t-1}, \mathbf{o}_t, \mathbf{h}_t)}
\big[
\log p_{\theta^{o}}(\mathbf{o}_t \mid \mathbf{z}_t, \mathbf{h}_t)
\big] \\
&-
D_{\mathrm{KL}}
\Big(
q_{\phi}(\mathbf{z}_t \mid \mathbf{z}_{t-1}, \mathbf{o}_t, \mathbf{h}_t)
\,\|\, 
p_{\theta^{z}}(\mathbf{z}_t \mid \mathbf{z}_{t-1}, \mathbf{h}_t)
\Big)
\Big].
\end{aligned}
\end{equation}
where $q_{\phi}(\mathbf{z}_t | \mathbf{z}_{t-1}, \mathbf{o}_t, \mathbf{h}_t)$ is the variational posterior that approximates the intractable true posterior. 
Specifically, we model $q_{\phi}(\mathbf{z}_t | \mathbf{z}_{t-1}, \mathbf{o}_t, \mathbf{h}_t)$ as a Gaussian distribution
\begin{equation}
q_{\phi}(\mathbf{z}_t \mid \mathbf{z}_{t-1}, \mathbf{o}_t, \mathbf{h}_t)
=
\mathcal{N}
\left(
\boldsymbol{\mu}^{\mathrm{post}}_t,
\boldsymbol{\Sigma}^{\mathrm{post}}_t
\right).
\end{equation}
The parameters $\boldsymbol{\mu}^{\text{post}}_t \in\mathbb{R}^{4\times1}$ and $\boldsymbol{\Sigma}^{\mathrm{post}}_t = \mathrm{diag}\!\left((\boldsymbol{\sigma}^{\text{post}}_t)^2\right)\in\mathbb{R}^{4\times4}$ are predicted by MLP networks with trainable parameters ${\phi}_\mu$ and ${\phi}_\sigma$, which together form the posterior net shown in Fig.~2. These networks map the inputs $\mathbf{z}_{t-1}, \mathbf{o}_t, \mathbf{h}_t$ to the corresponding distribution parameters
\begin{align}
\boldsymbol{\mu}^{\text{post}}_t
&=
f_{{\phi}_\mu}
(
\mathbf{z}_{t-1}, \mathbf{o}_t, \mathbf{h}_t
), \\
\boldsymbol{\sigma}^{\text{post}}_t
&=
\mathrm{Softplus}
\big(
f_{{\phi}_\sigma}
(\mathbf{z}_{t-1}, \mathbf{o}_t, \mathbf{h}_t)
\big)
+\epsilon_\Sigma,
\end{align}
where $\epsilon_\Sigma > 0$ is a small constant. During inference, the posterior mean is used as the state estimate $\hat{\mathbf{z}}_t = \boldsymbol{\mu}^{\text{post}}_t$.

% Given the generative model defined in Section III-B, the posterior distribution $p_{\theta}(\mathbf{z}_t \mid \mathbf{o}_{1:t})$ is generally intractable due to the nonlinear observation model and context-dependent uncertainty. Therefore, we introduce a variational distribution, denoted by $q_{\phi}(\mathbf{z}_t \mid \mathbf{o}_{1:t})$, to approximate the posterior induced by the generative model.

%\blue{resort to variational inference}\red{I don't understand too much about this one} to approximate the intractable posterior induced by the generative model.

% where the variational posterior is parameterized in a causal form:
% \begin{equation}
% q_{\phi}(\mathbf{z}_t \mid \mathbf{o}_{1:t})
% \approx
% q_{\phi}(\mathbf{z}_t \mid \mathbf{z}_{t-1}, \mathbf{o}_t, \mathbf{h}_t).
% \end{equation}

% This formulation follows a recursive filtering structure, where the posterior at time $t$ depends on the previous state, the current observation, and the temporal context.
%%%%%%%%

% This amortized inference can be interpreted as a learned approximation to the Bayesian filtering update, where the encoder implicitly integrates prior information from $\mathbf{z}_{t-1}$ and observation information from $\mathbf{o}_t$ under context-dependent uncertainty.
%{\color{red}use symbol where possible}
The term $\mathbb{E}_{q_{\phi}}[\log p_{\theta^{o}}(\cdot)]$ is the reconstruction loss, which corresponds to the expected log-likelihood. 
Specifically, it encourages the inferred latent state to be informative to explain the observations $\mathbf{o}_t$ through the observation model. 
The term $D_{\mathrm{KL}}(\cdot)$ is the Kullback–Leibler (KL) divergence between the variational posterior and the transition prior, which regularizes the posterior to remain consistent with the underlying state dynamics and prevents overfitting to noisy observations.
Together, these two terms balance observation fitting and dynamical consistency. 
% This is particularly important in WiFi-based localization, where observations are noisy, indirect, and exhibit time-varying reliability due to multipath propagation. 
As a result, the model achieves robust sequential inference under heterogeneous and non-stationary sensing conditions.

%{\color{green}provide a bit insight on why this works and link to the challenges of CSI localization}

\begin{figure}[ht]
    \centering
    \includegraphics[width=0.36\textwidth]{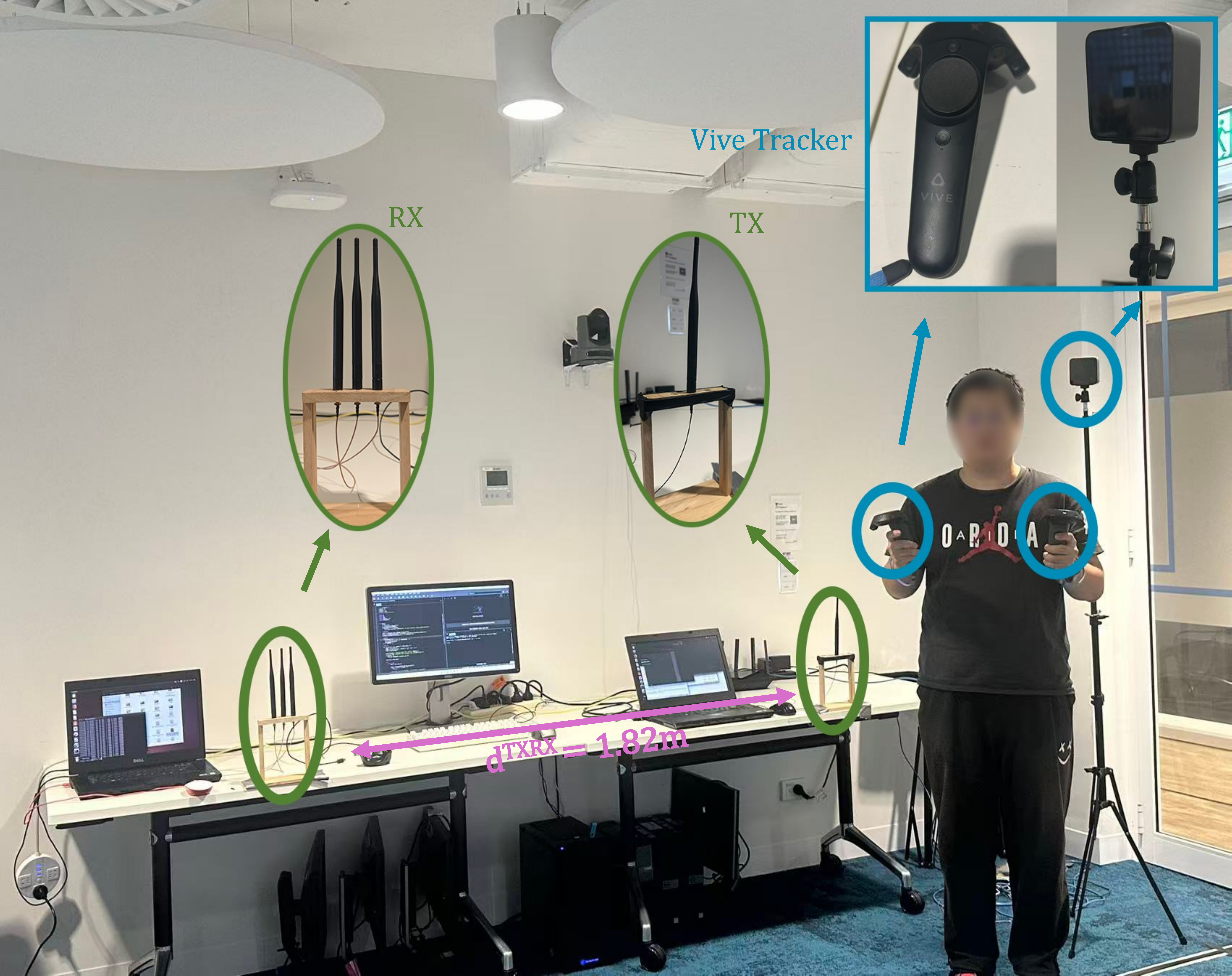}
    \vspace{-0.1cm}
    \caption{An illustration of the experimental environment.}
    \vspace{-0.6cm}
    \label{fig:env}
\end{figure}

\section{Experimental Results }
\subsection{Experimental Setup and Performance Metric}
        We evaluate the proposed system in a typical office setting, characterized by rich multipath propagation. As illustrated in Fig.~\ref{fig:env}, a transmitter with a single antenna and a receiver with three antennas $M=3$ are co-planarly deployed on a flat horizontal surface. The spacing between two adjacent antennas is $d=0.026$~m, and the transmitter--receiver distance is $d^{\mathrm{txrx}}=1.82$~m.
        The carrier frequency of the system $f^{\mathrm c}$ is 5.32~GHz with a 40~MHz channel bandwidth and $K=30$ different subcarriers. The sampling rate $f^{\rm s}$ of CSI is 1~kHz using Intel 5300 NICs in monitor mode~\cite{tools}. For AoA and ToF estimation, the window length $S$ is set to $100$. During inference, the human is walking at a velocity of around $1$m/s.
        %{\color{red}unit format inconsistent with above e.g., m and cm}. 
        The ground truth trajectories are obtained using a SteamVR-based Vive Tracker system with sub-centimeter accuracy~\cite{steamvr}.
        %\cite{steamvr}.
        The DSSM is trained with AdamW and a batch size of 32. The learning rate is $1\times10^{-3}$, the hidden state dimension is $d_h=128$, and the model is trained for 300 epochs with a cosine annealing schedule. 
        % The prior and posterior networks take inputs of dimensions 
        % $d_{\text{prior}}=132\,(4+128)$ and $d_{\text{post}}=135\,(4+3+128)$, respectively, 
        % and output 4-dimensional vectors; the observation-variance network outputs a 3-dimensional vector.
        %\red{also provide the dimension of prior/posterior and obs net}
        %The localization error at the $t$-th window is the Euclidean distance between the estimated location and the ground truth, defined as $\|\mathbf{p}_t-\hat{\mathbf{p}}_t\|$. 
        %The mean absolute error, median error, and cumulative distribution function (CDF) of the error of the trajectories can be calculated accordingly.
        %{\color{red}may delete if need space}. 
        
        %\red{check all the parameters above, i just give an erxample here}

\subsection{Performance Evaluation}       
We evaluate the proposed method against classical filtering approaches and advanced parameter estimation pipelines. Specifically, we compared the localization accuracy of combining SAGE~\cite{10.1145/3210240.3210314} and WiDFS~\cite{9679386} with EKF~\cite{barshalom2011tracking}, PF~\cite{gustafsson2010pf}, and the proposed DSSM-based framework. 
\begin{figure}[ht]
    \centering
    \vspace{-0.1cm}
    \begin{subfigure}{0.48\linewidth}
        \centering
        \includegraphics[width=\linewidth]{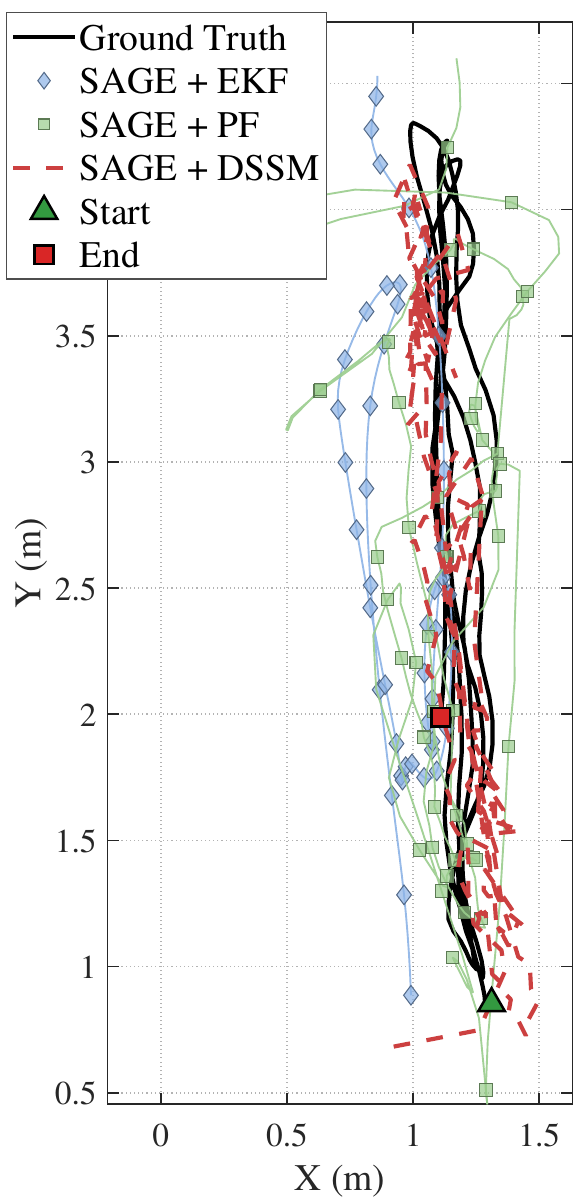}
        \caption{SAGE-based}
        \label{fig:sage}
    \end{subfigure}
    \hfill
    \begin{subfigure}{0.48\linewidth}
        \centering
        \includegraphics[width=\linewidth]{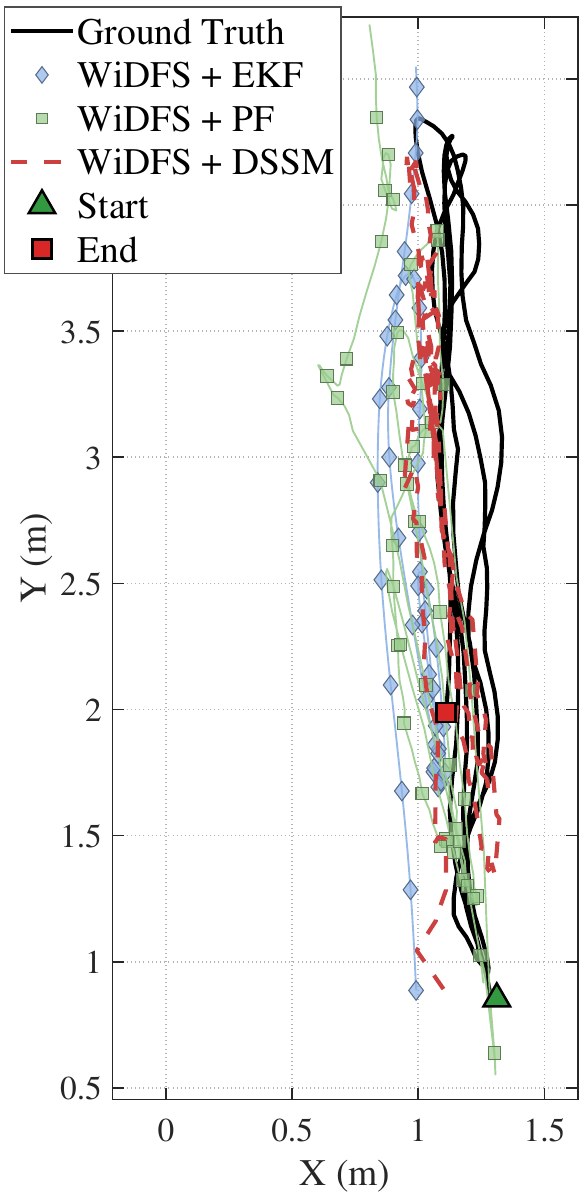}
        \caption{WiDFS-based}
        \label{fig:wdfs}
    \end{subfigure}
    % \vspace{-0.2cm}
    \caption{Trajectory comparison of different baseline methods.}
    \vspace{-0.3cm}
    \label{fig:traj_compare}
\end{figure}
Fig.~\ref{fig:traj_compare} shows the trajectories of the proposed framework and other baselines in both SAGE-based and WiDFS-based parameter estimation methods. The results indicate that the proposed framework closely follows the ground truth trajectory with smoother transitions, while EKF- and PF-based methods exhibit larger deviations and noticeable instability, especially under dynamic conditions.

The localization performance in terms of MAE, median error, 90\% cumulative distribution function (CDF) error, and standard deviation is summarized in Table~\ref{tab:performance}. The proposed DSSM-based methods achieve the lowest errors across all metrics. In particular, when combined with SAGE and WiDFS, the proposed DSSM reduces the MAE by 25.1\% and 31.9\%, respectively, compared with the corresponding raw methods. Similar improvements are observed in the median error, 90th percentile CDF, and standard deviation, demonstrating enhanced estimation stability.

We further evaluate the overall localization performance through the CDF of localization error in Fig.~\ref{fig:CDF}.
The proposed DSSM-based methods consistently outperform all baselines across the entire error range, achieving a higher probability of low-error estimates. For the proposed DSSM with WiDFS, 50\% and 90\% of the localization errors fall within 0.6~m and 1.3~m, respectively, compared to 0.77~m and 1.55~m for WiDFS with EKF, and 0.73~m and 1.8~m for WiDFS with PF. This indicates that the proposed method provides more reliable localization, particularly in the low-error regime.

{\setlength{\tabcolsep}{2.5pt}
\begin{table}[t]
\vspace{-0.0cm}
\centering
\caption{Performance comparison of different methods.}
\vspace{-0.1cm}
\label{tab:performance}
\begin{tabular}{lccccc}
\toprule
\textbf{Method} & \textbf{MAE (m)} & \textbf{Median (m)} & \textbf{P90 (m)} & \textbf{Std (m)} & \textbf{vs Raw} \\
\midrule

SAGE            & 0.978 & 0.777 & 1.934 & 0.694 & -- \\

\rowcolor{gray!20}
WiDFS           & 0.983 & 0.821 & 1.741 & 0.633 & -- \\

SAGE + EKF      & 0.852 & 0.695 & 1.558 & 0.568 & $\downarrow$12.9\% \\

\rowcolor{gray!20}
WiDFS + EKF     & 0.889 & 0.771 & 1.549 & 0.532 & $\downarrow$9.6\% \\

SAGE + PF       & 0.929 & 0.755 & 1.841 & 0.669 & $\downarrow$5.0\% \\

\rowcolor{gray!20}
WiDFS + PF      & 0.905 & 0.733 & 1.802 & 0.637 & $\downarrow$8.0\% \\

\midrule

\textbf{SAGE + DSSM}  & \textbf{0.733} & \textbf{0.570} & 1.493 & 0.564 & $\downarrow$25.1\% \\

\rowcolor{gray!20}
\textbf{WiDFS + DSSM} & \textbf{0.670} & \textbf{0.576} & \textbf{1.268} & \textbf{0.478} & $\downarrow$31.9\% \\

\bottomrule
\end{tabular}
% \vspace{-0.6cm}
\end{table}
}

\begin{figure*}[!t]
    \centering
    \vspace{-0cm}
    \begin{subfigure}[t]{0.32\textwidth}
        \centering
        \includegraphics[width=\linewidth]{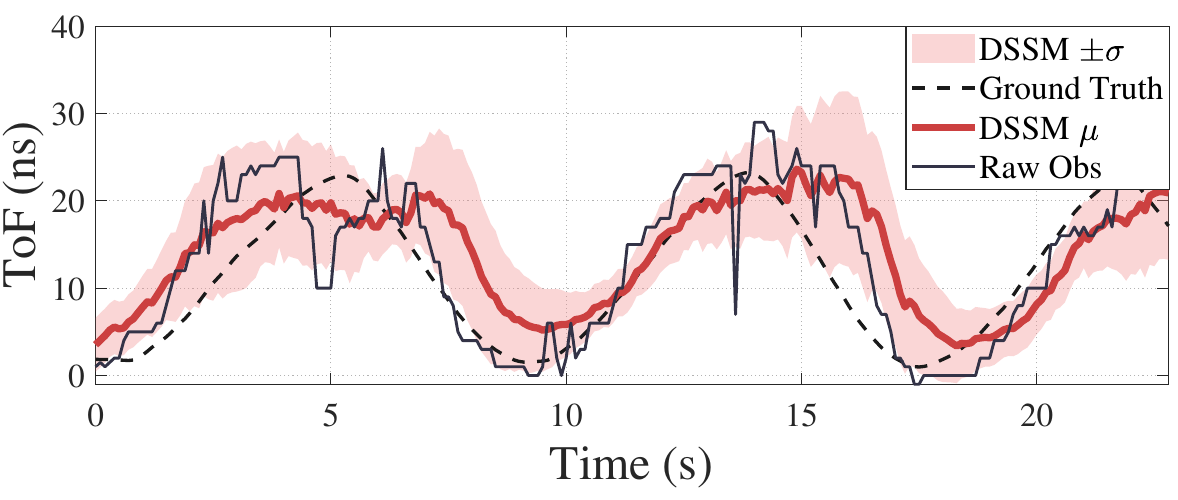}
        \caption{ToF comparison}
        \label{fig:tof_comparison}
    \end{subfigure}
    \hfill
    \begin{subfigure}[t]{0.32\textwidth}
        \centering
        \includegraphics[width=\linewidth]{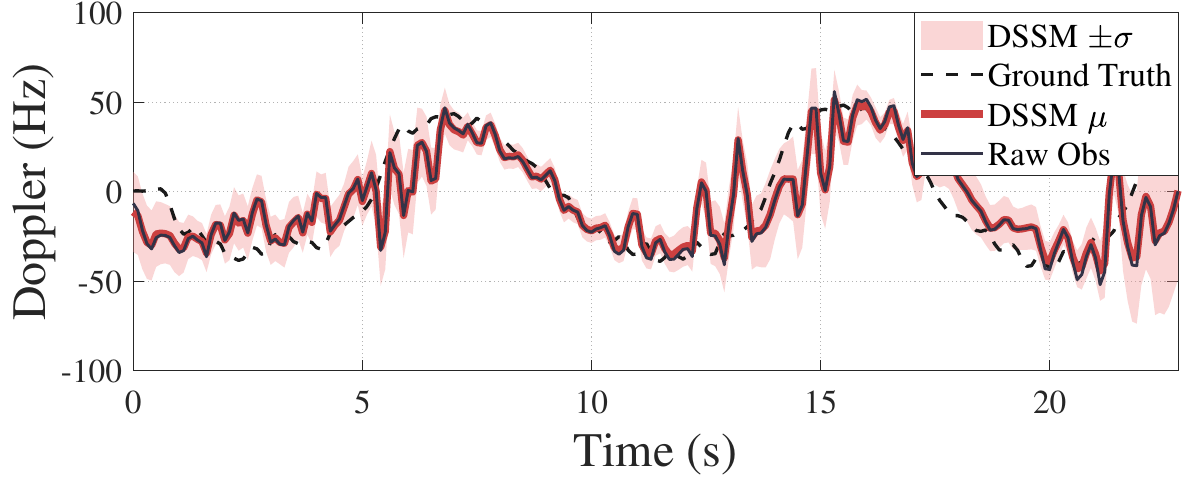}
        \caption{Doppler comparison}
        \label{fig:doppler_comparison}
    \end{subfigure}
    \hfill
    \begin{subfigure}[t]{0.32\textwidth}
        \centering
        \includegraphics[width=\linewidth]{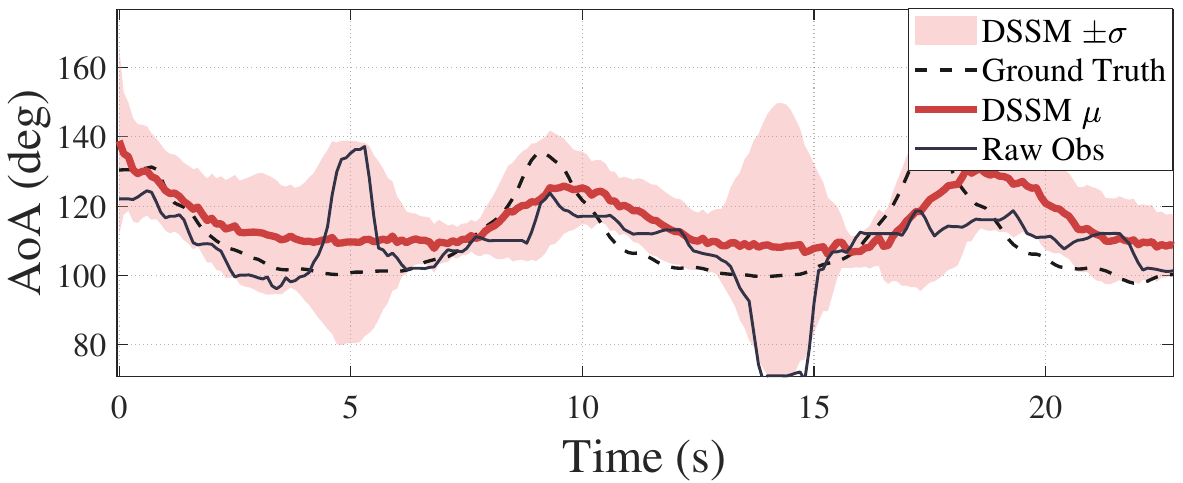}
        \caption{AoA comparison}
        \label{fig:aoa_comparison}
    \end{subfigure}
    \vspace{-0.15cm}
    \caption{Comparison of raw SAGE observations and DSSM estimates for ToF, Doppler, and AoA.}
    \label{fig:obs_distribution}
    \vspace{-0.35cm}
\end{figure*}
Fig.~\ref{fig:obs_distribution} shows the reconstructed observations of the AoA, ToF, and Doppler frequency shift of the SAGE method with the DSSM, respectively. The estimates are presented as mean values, $\boldsymbol{\mu}_t^{\mathrm{o}}$, and corresponding confidence bands, $\mathbf{R}_t$, together with the raw SAGE observations and the ground truth. It can be observed that the reconstructed ToF and Doppler closely follow the ground truth over time with significantly reduced fluctuations in the raw SAGE observations. 
For AoA, although the raw SAGE estimates exhibit noticeable abrupt variations, the proposed DSSM yields much smoother and more stable estimates that remain consistent with the overall trend of the ground truth. In addition, the shaded uncertainty bands vary over time, indicating that the proposed model can adaptively characterize the time-varying uncertainty of different observations.

%\begin{figure}[t]
%    \centering
 %   \includegraphics[width=0.50\textwidth]{figures/mat1.eps}
 %   \caption{Ablation study on different components.}
 %   \label{fig:ablation}
%\end{figure}
%\red{confused, how to remove the adaptive Q, R?, fixed q and R? how to work without physics? rewrite this part}
\begin{figure}[t]
    \centering
    \vspace{-0.2cm}
    \includegraphics[width=0.45\textwidth]{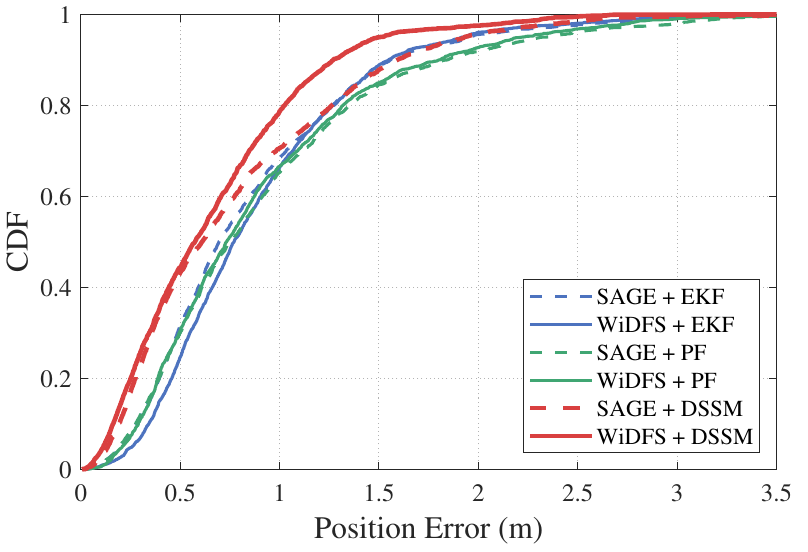}
    \vspace{-0.2cm}
    \caption{CDFs of localization errors for different methods.}
    \label{fig:CDF}
    \vspace{-0.5cm}
\end{figure}

To better understand the contribution of each component, Table~\ref {tab:ablation} reports the ablation results under different model variants. Specifically, \emph{w/o Context} replaces $\mathbf{h}_t$ with a learnable constant shared across all time steps, \emph{w/o Adaptive $\mathbf{Q},\mathbf{R}$} uses fixed diagonal covariance matrices, and \emph{w/o Physics} replaces the geometry-based mapping $\mathbf{g}(\mathbf{z}_t)$ with a generic multi-layer perceptron. As shown in Table.~\ref{tab:ablation}, the full model achieves the best performance with an MAE of 0.733\,m, compared with 1.421\,m, 1.115\,m, and 1.996\,m for the variants without context, adaptive uncertainty, and physics constraints, respectively, indicating the effectiveness of each component.
\begin{table}[ht]
\vspace{-0.4cm}
\centering
\setlength{\tabcolsep}{3pt}
\caption{Ablation study of the proposed framework.}
\vspace{-0.2cm}
\label{tab:ablation}
\begin{tabularx}{\columnwidth}{lccc|c}
\toprule
\textbf{Variant} 
& \textbf{Physics} 
& \textbf{Context ($h_t$)} 
& \textbf{Adaptive $Q, R$} 
& \textbf{MAE (m)} \\
\midrule
Full Model            & \checkmark & \checkmark & \checkmark & \textbf{0.733} \\
w/o Context           & \checkmark & $\times$   & \checkmark & 1.421 \\
w/o Adaptive $Q, R$   & \checkmark & \checkmark & $\times$   & 1.115 \\
w/o Physics           & $\times$   & \checkmark & \checkmark & 1.996 \\
\bottomrule
\end{tabularx}
\vspace{-0.4cm}
\end{table}

\section{Conclusion}
\vspace{-0.1cm}
In this paper, we propose a physics-informed DSSM for motion estimation from WiFi CSI. By modeling motion as a latent process and linking multi-domain measurements through geometric relationships, the method enables consistent integration of different sensing modalities.
A context-conditioned deep state space model was introduced to adapt transition and observation uncertainty over time. Experimental results demonstrate clear improvements in both accuracy and robustness over classical approaches, and ablation studies confirm the importance of each component.
The proposed framework provides an effective and practical solution for motion estimation from noisy CSI measurements.
% \vspace{-0.5cm}
\bibliographystyle{IEEEtran}
\bibliography{ref}
    
\end{document}